\documentclass{emulateapj}
\usepackage{apjfonts}

\slugcomment{To appear in the Astrophysical Journal}

\shorttitle{Proper motion of the Leo~II dwarf spheroidal galaxy}

\begin{document}

\title{A first measurement of the Proper Motion of the Leo~II dwarf
  spheroidal galaxy\altaffilmark{1}}

\author{S\'ebastien L\'epine\altaffilmark{2}, Andreas
  Koch\altaffilmark{3,4}, R. Michael Rich\altaffilmark{5}, and Konrad
  Kuijken\altaffilmark{6}}

\altaffiltext{1}{Based on observations made with the NASA/ESA Hubble
  Space Telescope, obtained from the Data Archive at the Space
  Telescope Science Institute, which is operated by the Association of
  Universities for Research in Astronomy, Inc., under NASA contract
  NAS 5-26555. These observations are associated with program
  \#11182.}

\altaffiltext{2}{Department of Astrophysics, Division of Physical
  Sciences, American Museum of Natural History, Central Park West at
  79th Street, New York, NY 10024, USA}

\altaffiltext{3}{Department of Physics and Astronomy, University of
  Leicester, University Road, Leicester LE1 7RH, UK}

\altaffiltext{4}{Zentrum f\"ur Astronomie der Universit\"at
  Heidelberg, Landessternwarte, K\"onigstuhl 12, 69117 Heidelberg,
  Germany}

\altaffiltext{5}{UCLA, Department of Physics and Astronomy, 430
  Portola Plaza, Los Angeles, CA 90095, USA}

\altaffiltext{6}{Leiden Observatory, Leiden University, PO Box 9513,
  2300RA Leiden, the Netherlands}

\begin{abstract}
We use 14-year baseline images obtained with the {\it Wide Field
Planetary Camera 2 } on board the Hubble Space telescope to derive a
proper motion for one of the Milky Way's most distant dwarf spheroidal
companions, Leo~II, relative to an extragalactic background reference
frame. Astrometric measurements are performed in the effective point
spread function (ePSF) formalism using our own developed code. An
astrometric reference grid is defined using 3,224 stars that are
members of Leo~II that are brighter than magnitude 25 in the F814W
band. We identify 17 compact extra-galactic sources, for which we
measure a systemic proper motion relative to this stellar reference
grid. We derive a proper motion
[$\mu_{\alpha}$,$\mu_{\delta}$]=[$+104\pm$113,-33$\pm$151] $\mu$as
yr$^{-1}$ for Leo~II in the heliocentric reference frame. Though
marginally detected, the proper motion yields constraints on the orbit
of Leo~II. Given a distance of d$\simeq$230 Kpc and a heliocentric
radial velocity $v_r=+79$ km s$^{-1}$, and after subtraction of the
solar motion, our measurement indicates a total orbital motion
$v_{G}=266.1\pm$128.7 km s$^{-1}$ in the Galactocentric reference
frame, with a radial component $v_{r_G}=21.5\pm$4.3 km s$^{-1}$ and
tangential component $v_{t_G}$=265.2$\pm$129.4 km s$^{-1}$. The small
radial component indicates that Leo~II either has a low-eccentricity
orbit, or is currently close to perigalacticon or apogalacticon
distance. We see evidence for systematic errors in the astrometry of
the extragalactic sources which, while close to being point sources,
are slightly resolved in the HST images. We argue that more extensive
observations at later epochs will be necessary to better constrain the
proper motion of Leo~II. We provide a detailed catalog of the stellar
and extragalactic sources identified in the HST data which should
provide a solid early-epoch reference for future astrometric
measurements.
\end{abstract}



\section{Introduction}

The Leo~II dwarf spheroidal (dSph) galaxy was originally discovered in
the course of the Palomar Sky Survey \citep{1950PASP...62..118H}. Its
great distance from the Milky Way (MW) marked it as being significant,
so it was the target of early studies \citep[e.g.][]{1962AJ.....67..125H}.
\citet{1995AJ....109..151V} found evidence for dark matter from its 11
km\,s$^{-1}$ velocity dispersion, while \citet{1996AJ....111..777M}
obtained the first HST-based color magnitude diagram (CMD) from WFPC2
photometry, deriving an age of $9\pm 1$ Gyr for its intermediate age
population. The most recent distance estimates, measured from
the tip of the red giant branch, place Leo~II at  $233\pm 15$ Kpc
\citep{2005MNRAS.360..185B} from the MW, and it remains as one of the
most distant known MW satellites that are candidates to be bound to
the Galaxy.  A recent analysis of 171 stellar radial velocities
by \citet{2007AJ....134..566K} finds no evidence that the galaxy is
experiencing the effects of tidal disruption, and indicates a mass to
light ratio of 25--50 (M/L)$_{\odot}$.  The Leo~II dwarf galaxy is of
great interest not only as a distant companion that offers a
constraint on the mass of the Milky Way
\citep[e.g.][]{2010MNRAS.406..264W}, but also as one of the
nearest galaxies that might have conceivably undergone evolution in
isolation from the effect of the Milky Way. Furthermore, its orbital
path provides an important feature of the evolutionary history of the
Leo~II dwarf.

Proper motions of other nearby dSphs have been detected and estimated
using images from the {\it Hubble Space Telescope} (HST). A
proper motion was determined for the Fornax dSph (d$\approx 140$ kpc)
by \citet{Dinescu2004},  \citet{Piatek2002}, and \citet{Piatek2007},
for Ursa Minor (d$\simeq$66 kpc) by \citet{Piatek2005}, Sculptor
(d$\simeq$79 kpc) by \citet{Piatek2006}, Carina (d$\simeq$
100 kpc) \citet{Piatek2003}, and the Small Magellanic Cloud
(d$\simeq$58 kpc) by \citet{Kallivayalil2006}. In all cases, the
orbital integration based on current Galactic mass models yields bound
orbits, with apogalactic distances not exceeding $\sim$150 kpc,
although including the proper motion uncertainties, one cannot rule
out apogalactic
distances as high as 300 kpc for Sculptor at the 95\% confidence level
\citep{Piatek2006}. However, recent proper motion measurements of the
Large and Small Magellanic Clouds yield space velocities larger than
expected, which suggests that the two objects may be only marginally
bound to the Galaxy \citep{Kallivayalil2009}.

Due to their relatively large distances, the proper motions of dSphs
are at the sub-pixel level over the baselines typically available from
HST archival images ($<20$ years). For example, at a distance of
100 kpc, a 100 km\,s$^{-1}$ velocity translates into a proper motion of only
0.211 mas yr$^{-1}$, which yields a net motion of only 0.03 pixels on
the WFPC2 camera over a period of 15 years. Due to the undersampling
of the point spread function in HST images, the detection of such a
small astrometric motion requires a special astrometric reduction
procedure. The {\em effective Point Spread Function} (ePSF) method,
pioneered by \citet{AK2000} and \citet{Piatek2002}, is well suited to
the task. 

We have developed our own reduction software, based on the ePSF
method, in order to detect and calculate the proper motion of Leo~II.
This paper presents initial results from this analysis,
which allows us for the first time to investigate whether this remote
dSph is an actual, bound satellite to the MW and, if so, to study in
detail its orbital characteristics. The archival HST datasets used in
our study are described in \S2. The astrometric reduction is
summarized in \S3. The identification of extragalactic sources,
critical in establishing an astrometric reference frame, is
described in \S4. The proper motion of Leo~II is calculated in
\S5. The inferred space motion and orbit of Leo~II is discussed in
\S6.

\section{HST observations and archival data}

The dSph Leo~II was  first observed with the Wide Field and
Planetary Camera 2 (WFPC2) on May 15, 1994. In addition to a pair of
shallow 80s exposures in the F555W band, a total of eight deep
exposures (600 seconds each) were obtained in each of the F555W and
F814W bands. Dithering patterns were not generally used at the time,
and the 1994 frames have no significant offsets between them.

Leo~II was reobserved on March 19, 2004. This time, only eight
exposures in the F814W band were obtained, 500-700 seconds each. The 2004
fields are aligned with the 1994 fields to within $\approx3\arcsec$,
and with the same field orientation. The 2004 exposures followed a
4-point dithering pattern. Frames 1 and 2 constitute a pair with no
offset between each other, as are the pairs consisting of frames 3-4,
5-6, and 7-8. The 4-point dithering pattern occurs between each of the
pairs.

Leo~II was observed again on March 25, 2008, yielding eight exposures
of 1,100 seconds in each of the F555W and F814W bands. This time an 8-point
dithering pattern was used, so no two images are perfectly aligned,
and display small offsets between each other. The 2008 frames have the
same orientation as the 1994 and 2004 frames, but are offset by
20$\arcsec$ in the direction of Right Ascension. This means that on
one side of each WF camera frame from 1994/2004, there is a band
$\approx200$ pixels wide which is not covered by the 2008
exposures. This leaves an area approximately 4 arcmin$^2$, or 75\% of
the field of view of the Wide Field Camera chips, which was imaged at
all three epochs. The center of the WFPC2 field of view was offset
only $\approx1\arcmin$ from the center of the dwarf spheroidal. With
Leo~II having a core radius of 2.64$\arcmin$
\citep{2007AJ....134.1938C}, the HST observations thus cover
$\sim 22\%$ of the area within the core radius of Leo~II.

\begin{figure}
\epsscale{1.15}
\plotone{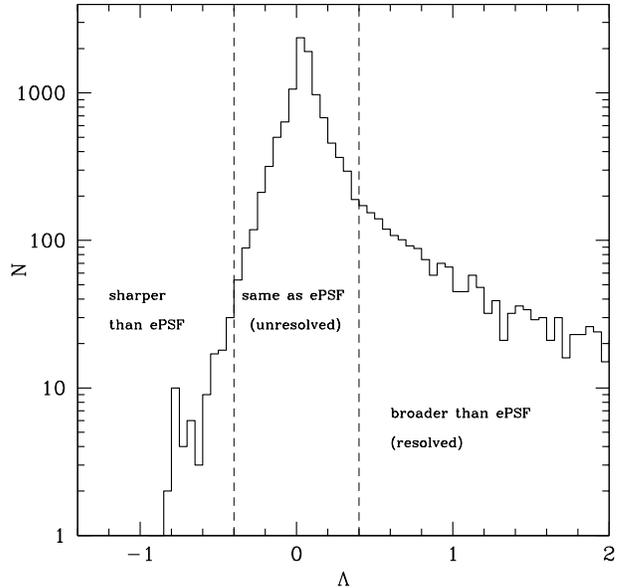}
\caption{Statistical distribution of the second moment of the
  distribution $\Lambda$ (or ``sharpness parameter'') for sources
  detected in the Leo~II field by the WFPC2 camera. A value of
  $\Lambda\approx0$ indicates that the source has a spread similar to
  the local ePSF, which means it is unresolved by the camera. The
  distribution shows a tail of objets with $\Lambda>1.4$; these are
  sources which are effectively resolved by the camera.}
\end{figure}

\section{Astrometric analysis}

\begin{deluxetable*}{ccccccccc}
\tabletypesize{\scriptsize}
\tablecolumns{9} 
\tablewidth{0pt} 
\tablecaption{Stellar members of Leo~II defining the astrometric grid\tablenotemark{a}} 
\tablehead{ 
\colhead{$\alpha$} & 
\colhead{$\delta$} & 
\colhead{$\mu_{\alpha}$\tablenotemark{b}} & 
\colhead{$\mu_{\delta}$\tablenotemark{b}} & 
\colhead{F555W} & 
\colhead{F814W} & 
\colhead{F555W-F814W} & 
\colhead{$\Lambda$} &
\colhead{camera} 
\\
\colhead{(J2000)} & 
\colhead{(J2000)} & 
\colhead{(mas yr$^{-1}$)} & 
\colhead{(mas yr$^{-1}$)} & 
\colhead{(mag)} & 
\colhead{(mag)} & 
\colhead{(mag)} & 
\colhead{} &
\colhead{WF...}
}
\startdata
168.3551962$\pm$0.0000007&  +22.1485758$\pm$0.0000007&  -0.34$\pm$0.36&   0.97$\pm$0.37&  25.90&  25.26&  0.63&  0.17 & 2\\
168.3552438$\pm$0.0000009&  +22.1480854$\pm$0.0000009&   0.51$\pm$0.36&   0.12$\pm$0.49&  25.98&  25.41&  0.56& -0.29 & 2\\
168.3553641$\pm$0.0000001&  +22.1512710$\pm$0.0000002&   0.07$\pm$0.06&   0.05$\pm$0.10&  22.84&  21.93&  0.91&  0.09 & 2\\
168.3553896$\pm$0.0000006&  +22.1492520$\pm$0.0000006&   0.71$\pm$0.26&  -0.79$\pm$0.33&  26.02&  25.59&  0.43&  0.01 & 2\\
168.3553999$\pm$0.0000002&  +22.1519683$\pm$0.0000004&   0.47$\pm$0.14&   0.07$\pm$0.22&  25.17&  24.71&  0.46&  0.14 & 2\\
168.3554060$\pm$0.0000003&  +22.1473196$\pm$0.0000002&   0.11$\pm$0.10&   0.01$\pm$0.10&  24.52&  23.73&  0.80&  0.05 & 2\\
168.3554769$\pm$0.0000006&  +22.1465200$\pm$0.0000004&  -0.03$\pm$0.21&   0.34$\pm$0.21&  25.38&  24.84&  0.55&  0.20 & 2\\
168.3555478$\pm$0.0000007&  +22.1454925$\pm$0.0000005&   0.01$\pm$0.29&   0.52$\pm$0.24&  25.39&  24.77&  0.63&  0.08 & 2\\
168.3555627$\pm$0.0000005&  +22.1502803$\pm$0.0000006&  -0.41$\pm$0.22&   0.26$\pm$0.29&  25.45&  24.83&  0.62&  0.02 & 2\\
168.3555798$\pm$0.0000009&  +22.1513628$\pm$0.0000005&   1.03$\pm$0.33&   0.69$\pm$0.28&  25.80&  25.20&  0.60&  0.03 & 2\\
168.3556039$\pm$0.0000003&  +22.1465930$\pm$0.0000002&   0.14$\pm$0.10&  -0.12$\pm$0.11&  23.88&  23.05&  0.82&  0.19 & 2\\
168.3556093$\pm$0.0000008&  +22.1532404$\pm$0.0000004&   0.30$\pm$0.22&   0.08$\pm$0.17&  25.84&  25.30&  0.54& -0.13 & 2\\
168.3556196$\pm$0.0000009&  +22.1484238$\pm$0.0000004&   0.48$\pm$0.29&  -0.13$\pm$0.23&  25.08&  24.55&  0.53&  0.16 & 2\\
168.3556291$\pm$0.0000003&  +22.1536420$\pm$0.0000005&   0.01$\pm$0.15&  -1.01$\pm$0.24&  25.61&  25.06&  0.56&  0.10 & 2\\
168.3557186$\pm$0.0000010&  +22.1485511$\pm$0.0000008&  -0.29$\pm$0.44&  -0.75$\pm$0.38&  25.79&  25.15&  0.64&  0.26 & 2\\
168.3557224$\pm$0.0000005&  +22.1542072$\pm$0.0000004&  -0.21$\pm$0.23&  -0.40$\pm$0.22&  25.79&  25.32&  0.47&  0.20 & 2\\
168.3557244$\pm$0.0000002&  +22.1532839$\pm$0.0000006&   0.41$\pm$0.13&  -0.41$\pm$0.30&  25.26&  24.75&  0.51&  0.09 & 2\\
168.3557257$\pm$0.0000005&  +22.1510173$\pm$0.0000005&  -0.34$\pm$0.20&   0.59$\pm$0.25&  25.83&  25.15&  0.68& -0.07 & 2\\
168.3557325$\pm$0.0000003&  +22.1573483$\pm$0.0000005&   0.67$\pm$0.14&   0.08$\pm$0.28&  25.96&  25.49&  0.47&  0.18 & 2\\
168.3557424$\pm$0.0000049&  +22.1484209$\pm$0.0000009&   0.18$\pm$0.82&  -0.24$\pm$0.62&  26.31&  25.93&  0.38&  0.14 & 2\\
\enddata
\tablenotetext{a}{This table is available in its entirety in the
  electronic version of this paper published in the Astrophysical
  Journal website. The first 20 lines of the table are shown here as
  a guide to the table format.}
\tablenotetext{b}{Relative proper motions in the defined Leo~II
  stellar grid.}
\end{deluxetable*}

\begin{figure}
\epsscale{2.35}
\plotone{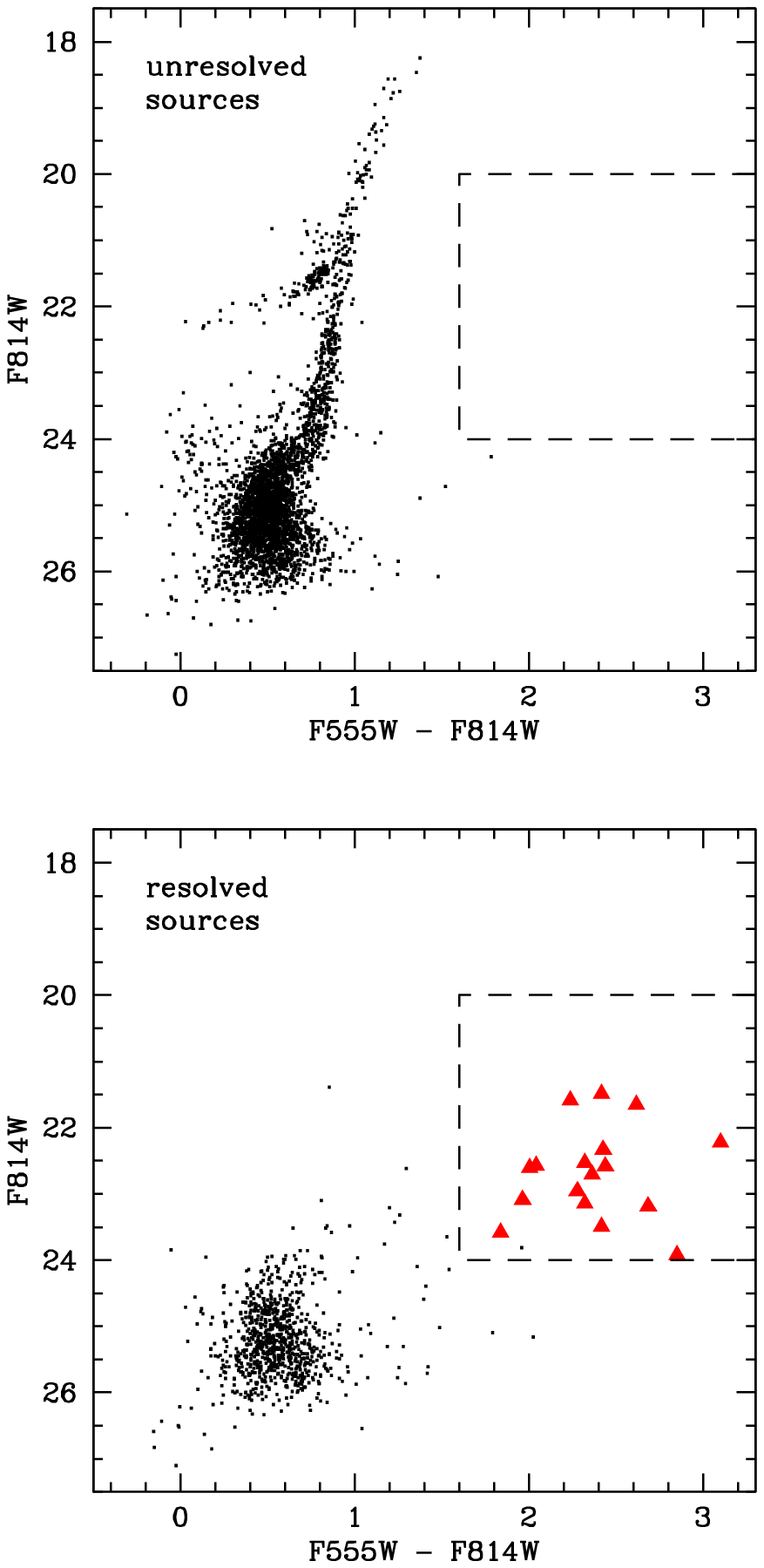}
\caption{Color magnitude diagram of the Leo~II field. All stars
  identified and measured with our astrometric code are shown. The
  upper panel shows the stars which remain unresolved after the ePSF
  fit. The lower panel shows all of the resolved objects (sources with a
  sharpness factor $\Lambda>$1.3). All of the unresolved
  sources produce a CMD consistent with Leo~II membership. The resolved
  sources fall along two distinct loci: they are either coincident
  with the low-mass stellar locus of Leo~II and hence are likely
  visual binaries, or they have very red colors, which then strongly
  suggest that they are extended extra-galactic sources
  (galaxies/quasars). Objects enclosed in the dashed-line box are
  formally identified as extragalactic sources. One faint source
  within the extragalatic cut (black dot within the dashed area) was
  rejected because of a large astrometric uncertainty.}
\end{figure}

The positions of all sources on the frames were re-calculated in the
effective point spread function (ePSF) formalism, using a method
analogous to the one introduced and developed by \citet{AK2000}
(hereafter AK) and \citet{Piatek2002}. The method consists of fitting
the pixel profile of a source $[P_k]_{i,j}=P_k(i,j)$ with a local
representation of the response of the camera to a point source
$\Psi_k=\Psi(i-x_k,j-y_k)$, where $(x_k,y_k)$ denote the
hypothetical location of the point source in the $(i,j)$ pixel
grid. Instead of being modeled by a mathematical function, the
functional form of $\Psi$, the ePSF, is extracted from the data using
the multiple samplings provided by the large number of point sources
in the field, because each of the $k$ sources provides samplings of
$\Psi$ at the locations $(i-x_k,j-y_k)$. Because the
hypothetical locations of the point sources $(x_k,y_k)$ are not known
{\it a priori}, the functional form of $\Psi$ must be determined using
an iterative procedure, where the $(x_k,y_k)$ are re-calculated and
the functional form of $\Psi$ re-evaluated after each iteration, as
described in \citep{AK2000}. We have developed our own reduction
software, which determines the ePSF directly out of the data, using an
iteration procedure, exactly as prescribed in AK. Our ePSF is
determined separately for each of the three epochs, and independently
for each band. An ePSF is also calculated separately for each of the
Wide Field Camera frames (WF2, WF3, WF4), and the ePSF is allowed to
vary continuously across the frame, following its separate
determination in each of nine sectors across the chip. Since our
general procedure is in most points identical to the one described in
\cite{AK2000}, we do not repeat the details here.

\begin{figure*}
\includegraphics[width=1\hsize]{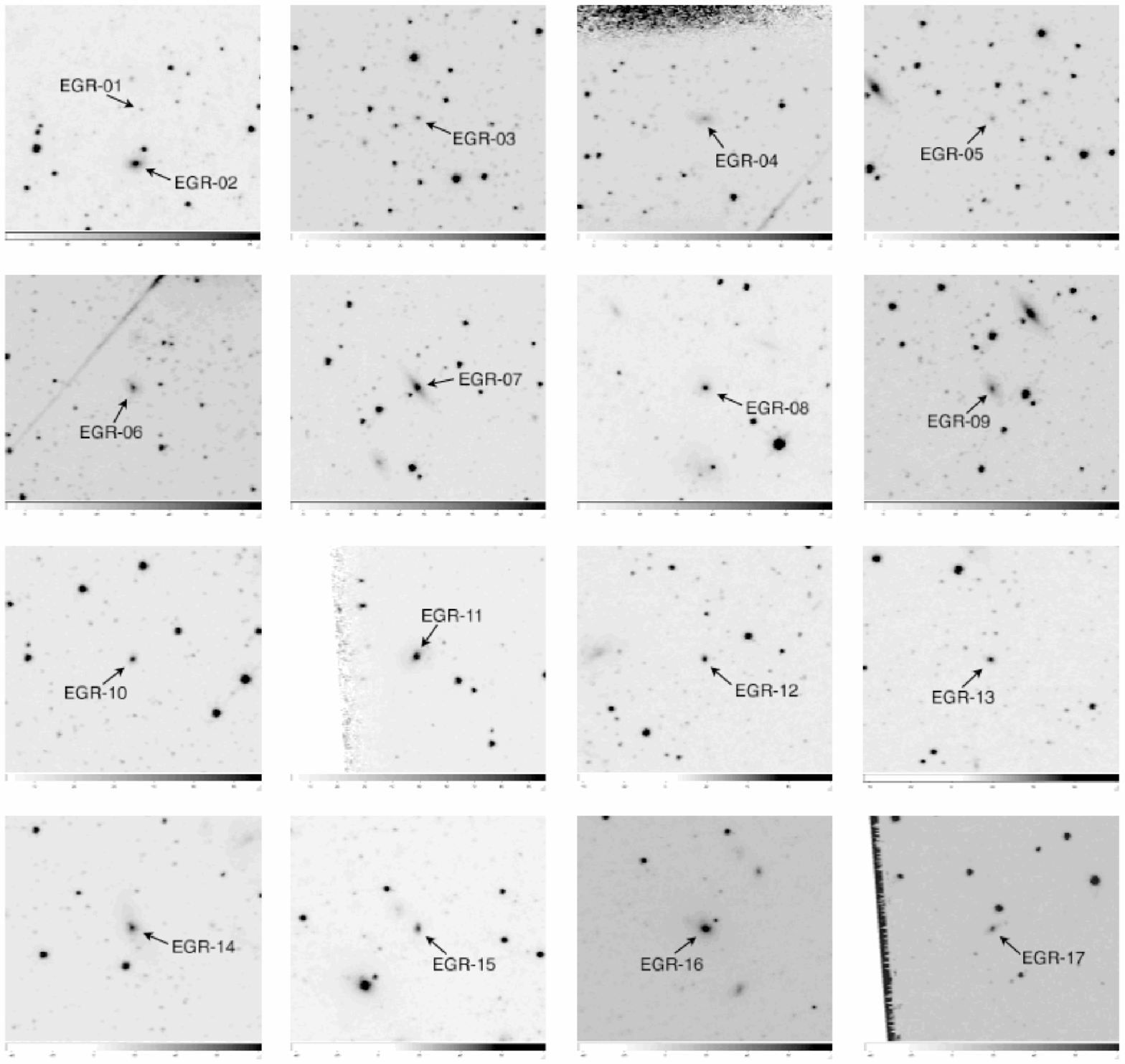}
\caption{Finder charts for the extra-galactic sources used as
  astrometric reference objects. All fields are 14.5$\arcsec$ on the
  side with North up and East left. Most sources are associated with
  an extended diffuse emission which identifies them as background
  galaxies.}
\end{figure*}

Once a satisfactory ePSF is determined, an algorithm is used to
converge the centroid $(x_k,y_k)$ of each source. The algorithm
differs from a least-square fit in that it uses a weighting of all the
pixels proportial to the local first derivative of the ePSF profile,
which gives more weight to the profile edges than to the central pixel
which often yield little positional information. The flux $f_k$ of the
source is re-evaluated after each iteration from an ePSF fit of the
stellar profile calculated for the current $(x_k,y_k)$
estimate. Convergence is attained after 5-10 iterations. A $\chi^2$
test is then used to evaluate the goodness of fit; sources which
differ significantly from the ePSF  \--- such as camera artifacts, cosmic ray
events, and very extended sources (e.g. galaxies) \--- are rejected. The
first moments of the residuals $\Gamma_k$ are calculated as:
\begin{equation}
\Gamma_k = \sum_{i,j} [P_k]_{i,j} - f_k \Psi(i-x_k,j-y_k) ,
\end{equation}
and should be $\approx$0 assuming $f_k$ is a close estimate of the
flux. Second moments of the distribution $\Lambda_k$ are then
calculated, which offer a quantification of how much intrinsic
``spread'' there is in the source compared to the local ePSF:
\begin{equation}
\Lambda_k = \sum_{i,j} \sqrt{(i-x_k)^2+(j-y_k)^2} \left( [P_k]_{i,j} - f_k
\Psi(i-x_k,j-y_k) \right) .
\end{equation}
The second moment is expected to be $\approx0$ only if the source has
approximately the same spread as the local ePSF --- i.e. it is
unresolved by the camera. But if the source has a profile different
from the ePSF, then $\Lambda_k$ could be significantly different from
0. This value $\Lambda_k$ thus works like a ``sharpness parameter'',
distinguishing between resolved and unresolved sources. We find that
the vast majority of the stars in the Leo~II field have
$-0.4<\Lambda_k<0.4$, which determines the range over which sources
can be considered point-like; we use those sources to define the local
astrometric grid. Sources with a second moment $\Lambda_k>0.4$, have a
significantly larger spread than expected from the local ePSF, and are
effectively resolved by the camera. This is clear in Fig.1 where we
see a tail of sources with larger $\Lambda_k$ values. Visual
examination of sources with large measured values of $\Lambda_k$
indeed reveals the sources to be noticeably extended compared with
most other sources in the field. Sources with a very large sharpness
factor ($\Lambda>3$) are so broadened compare to the ePSF that they
generally {\em fail} the $\chi^2$ test and are rejected by the code;
hence the sources which are detected by the code but flagged as
``resolved'' are still relatively compact objects.

The ``unresolved'' sources are used to build a relative astrometric
frame. Astrometric corrections are performed in multiple steps, in
order to bring the positions measured in all exposures/epochs into a
single common reference grid. Each unresolved source is used in
defining the grid with a weight proportional to its estimated
astrometric uncertainty; in effect this yields a larger weight on the
brighter sources. Corrections are performed first on individual frames
of a given epoch/band; these correct for small offsets, rotations, or
dilations between successive images, notably the offsets from the
dithering patterns and also changes in the field scale due to thermal
dilation in HST and the WFPC2. Average positions are then recalculated
for each epoch/band using the mean position of all the measurements
from that particular set. Additional corrections are then performed to
bring the calculated mean positions from all seven epochs/bands into a
common master grid. Finally, proper motions are calculated for each
individual source, by a linear regression of the mean position of the
source as a function of epoch.

Astrometric and photometric data for all of the unresolved (point-like)
sources is provided in Table 1. The table lists the measured positions
and relative proper motions of the 3,224 sources, along with F555W and F814W
magnitudes. We also list the second moments $\Lambda$ of the source
profiles. The WF camera frame (2-4) in which the source was
detected is also listed. These stars define our astrometric reference
grid.

\section{Identification of extra-galactic reference sources}

\begin{deluxetable*}{cccccccccc}
\tabletypesize{\scriptsize}
\tablecolumns{10} 
\tablewidth{0pt} 
\tablecaption{Extra-galactic reference sources in the Leo-II field}
\tablehead{ 
\colhead{ID} & 
\colhead{$\alpha$} & 
\colhead{$\delta$} & 
\colhead{$\mu_{\alpha}$\tablenotemark{a}} & 
\colhead{$\mu_{\delta}$\tablenotemark{a}} & 
\colhead{F555W} & 
\colhead{F814W} & 
\colhead{F555W-F814W} & 
\colhead{$\Lambda$} &
\colhead{camera} 
\\
\colhead{} & 
\colhead{(J2000)} & 
\colhead{(J2000)} & 
\colhead{(mas yr$^{-1}$)} & 
\colhead{(mas yr$^{-1}$)} & 
\colhead{(mag)} & 
\colhead{(mag)} & 
\colhead{(mag)} & 
\colhead{} &
\colhead{WF...}
}
\startdata
 EGR-01& 168.3577273$\pm$0.0000024&  +22.1751675$\pm$0.0000007&   0.36$\pm$0.82&  -0.29$\pm$0.37&  25.33&  22.23&  3.10&  0.64&  3\\
 EGR-02& 168.3578068$\pm$0.0000018&  +22.1766783$\pm$0.0000010&  -0.03$\pm$0.70&   0.48$\pm$0.51&  26.78&  23.92&  2.85&  0.73&  3\\
 EGR-03& 168.3625705$\pm$0.0000007&  +22.1525700$\pm$0.0000004&   0.52$\pm$0.24&  -0.06$\pm$0.24&  25.47&  23.15&  2.32&  0.53&  2\\
 EGR-04& 168.3629439$\pm$0.0000015&  +22.1628222$\pm$0.0000011&  -0.90$\pm$0.66&  -0.32$\pm$0.60&  25.41&  23.58&  1.84&  1.03&  2\\
 EGR-05& 168.3635967$\pm$0.0000008&  +22.1521349$\pm$0.0000004&   0.53$\pm$0.26&  -0.65$\pm$0.24&  25.91&  23.49&  2.42&  0.51&  2\\
 EGR-06& 168.3650347$\pm$0.0000011&  +22.1560831$\pm$0.0000009&   0.78$\pm$0.49&   1.08$\pm$0.49&  25.24&  22.96&  2.28&  0.73&  2\\
 EGR-07& 168.3658434$\pm$0.0000004&  +22.1526715$\pm$0.0000007&   1.09$\pm$0.32&   0.37$\pm$0.39&  24.27&  21.65&  2.62&  0.59&  2\\
 EGR-08& 168.3667410$\pm$0.0000007&  +22.1786564$\pm$0.0000006&  -0.25$\pm$0.39&   0.73$\pm$0.28&  24.76&  22.33&  2.43&  0.46&  3\\
 EGR-09& 168.3665803$\pm$0.0000012&  +22.1513390$\pm$0.0000010&  -1.13$\pm$0.50&   0.14$\pm$0.52&  25.87&  23.18&  2.69&  1.10&  2\\
 EGR-10& 168.3684006$\pm$0.0000003&  +22.1693588$\pm$0.0000008&  -0.46$\pm$0.26&  -0.10$\pm$0.51&  25.02&  22.58&  2.44&  0.75&  3\\
 EGR-11& 168.3718307$\pm$0.0000008&  +22.1709446$\pm$0.0000005&  -0.49$\pm$0.36&  -0.62$\pm$0.30&  23.82&  21.59&  2.24&  0.42&  3\\
 EGR-12& 168.3815968$\pm$0.0000001&  +22.1718691$\pm$0.0000009&  -0.38$\pm$0.12&  -0.87$\pm$0.47&  25.07&  22.70&  2.37&  0.48&  4\\
 EGR-13& 168.3853748$\pm$0.0000007&  +22.1681213$\pm$0.0000006&   0.71$\pm$0.34&   0.68$\pm$0.34&  24.61&  22.61&  2.01&  0.46&  4\\
 EGR-14& 168.3858761$\pm$0.0000006&  +22.1704062$\pm$0.0000014&  -1.23$\pm$0.47&  -1.53$\pm$0.75&  24.62&  22.57&  2.04&  1.05&  4\\
 EGR-15& 168.3900591$\pm$0.0000008&  +22.1725934$\pm$0.0000010&   1.47$\pm$0.46&   0.70$\pm$0.55&  25.06&  23.09&  1.97&  0.91&  4\\
 EGR-16& 168.3910732$\pm$0.0000003&  +22.1715817$\pm$0.0000004&  -1.08$\pm$0.16&   0.77$\pm$0.29&  23.90&  21.48&  2.42&  0.56&  4\\
 EGR-17& 168.3927718$\pm$0.0000006&  +22.1666944$\pm$0.0000006&  -0.18$\pm$0.24&  -0.42$\pm$0.42&  24.86&  22.53&  2.32&  0.47&  4\\
\enddata
\tablenotetext{a}{Relative proper motions in the defined Leo~II
  stellar grid.}
\end{deluxetable*}

A CMD of all of the sources in the combined Leo~II field is shown in
Fig.~2. Distinct plots are presented for the resolved (bottom panel)
and unresolved (top panel) sources, as defined in Fig.~1. The CMD
for the unresolved sources clearly displays the asymptotic giant
branch, horizontal branch, red giant sequence, and main sequence
expected for the dSph \citep[see
  also][]{1996AJ....111..777M,2007AJ....134.1938C,2007AJ....134..835K}. A
distinct population of blue stragglers is also detected. The CMD of
the resolved sources tells another story. The giant sequences are
indistinguishable. However, a significant number of objects are
detected on or near the main-sequence locus. Because their colors and
magnitudes are consistent with main sequence stars, we conclude that
these ``resolved'' sources are most likely visual binaries, probably
due to crowding in the dense Leo~II field.

The CMD of unresolved sources also displays a distinct
population of very red objects (F555W$-$F814W$>$1.4 mag). Many of
these are moderately bright and within the range ($V<23.5$) from
which we can expect accurate astrometry. Since no comparable sources
are detected in the unresolved population, we conclude that these are,
in fact, extended objects, with point spread functions that are
locally wider that the fiducial ePSF. This, combined with the fact
that they are found significantly off the main stellar locus, suggests
that they are background galaxies. A close examination of the sources
confirms this impression (Fig.~3).

We use the CMD to separate probable Leo~II members from
extragalactic background sources. The idea is that most extragalactic
objects should fall off the stellar locus. Distant galaxies in
particular should be redder than stellar members as a result of their
intrinsically red populations and redshift. To help in the
identification of an extragalactic reference set,
we particularly consider the distribution of resolved sources, which
are likely to include most of the background galaxies (see above). The
distribution of resolved source in Fig.~2 is unambiguous. A
significant number of resolved objects define a locus well to the red
of the {fiducial} stellar locus of Leo~II members. Sources with
F555W$-$F814W$>$1.3 and F555W$<$25.0 are indeed very unlikely to be stellar
members of Leo~II. We formally identify as extra-galactic sources all
those which fall inside the {empirical} box plotted in Fig.~2, except
for one faint source with a very large astrometric uncertaintry.

\begin{figure}
\epsscale{1.25}
\plotone{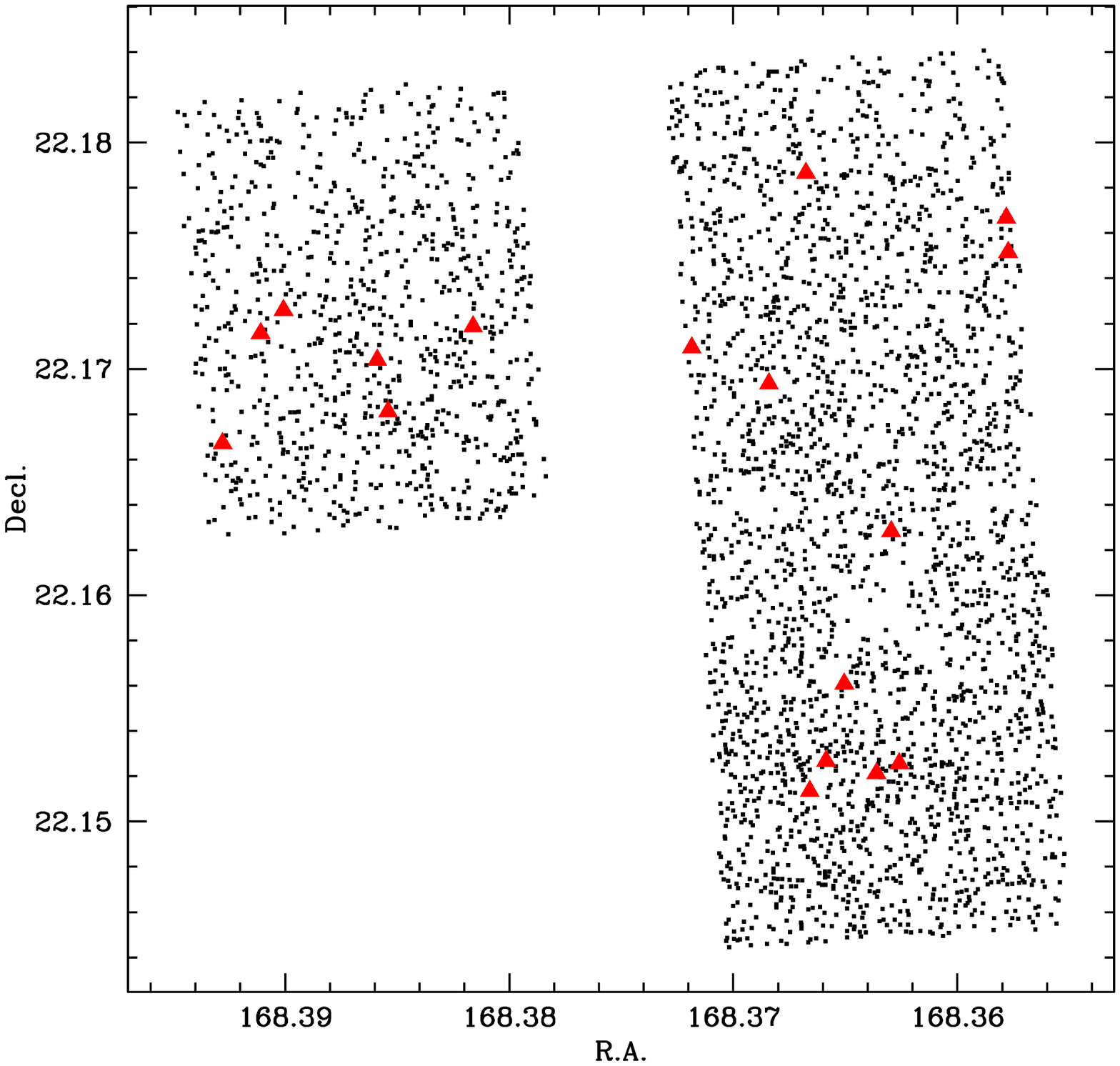}
\caption{Position of the astrometric reference stars from Leo~II
  (black dots) and reference extra-galactic sources (red
  triangles). Extra-galactic sources are found in each of the WFPC2 wide
  field camera frames. Note the rectangular shape of the fields which
  is due to an offset in the 2008 epoch images. The surveyed area
  covers 0.0011 square degrees near the center of Leo~II, representing
  $\approx$22\% of the area within the galaxy's core radius.}
\end{figure}

The resolved (Leo~II) sources, which define the local astrometric grid,
and the extra-galactic sources, which are used as background
astrometric reference, are all plotted in Fig.~4 as a function of
R.A. and declination. Extragalactic reference objects are distributed over
all three independent camera frames. The extragalactic sources are
also relatively spread out and sample both the centers and edges of
each field.

Astrometric and photometric data for the extra-galactic reference
objects is provided in Table 2. The table lists the measured positions
and proper motions of the 17 sources, along with F555W and F814W
magnitudes. We also list the second moments $\Lambda$ of the source
profiles (see \S3 above), which measures how the profiles compare to a
point source. The WF camera frame (2-4) in which the source was
detected is also listed.


\section{Relative proper motion of Leo~II members}

\begin{figure*}
\epsscale{1.0}
\plotone{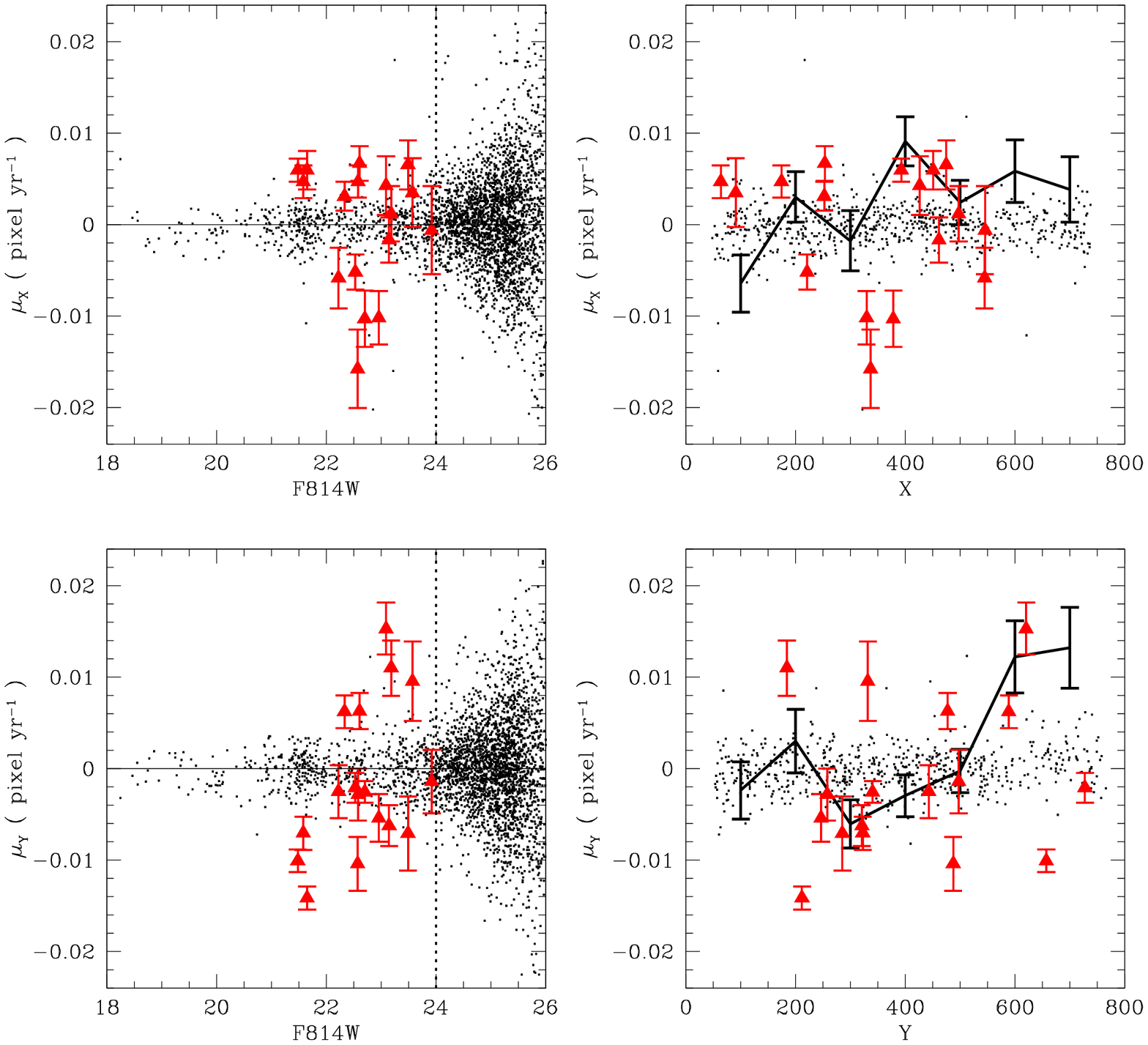}
\caption{Proper motions $\mu_X$ and $\mu_Y$ of the Leo~II field
  sources along the camera X and Y coordinates, for all three WF
  cameras. Unresolved stellar sources are plotted as black dots, while
  the extra-galactic reference objects as plotted as red triangles (with
  errorbars). Panels on the left show proper motions as a function of
  the source magnitude in the F814W band. The stability of the
  reference frame allows for proper motion measurements with a mean accuracy
  $\pm4$ millipixels per year for unresolved sources brighter than
  F814W=24, but resolved extragalactic sources show a noticably larger
  scatter. Panels on the right show $\mu_X$ and $\mu_Y$ as a function
  of column number X and row number Y, respectively. Individual points
  are shown for bright sources (F814W$<$24.0, both stellar and
  extra-galactic) while the mean values are shown for the fainter
  sources (broken line) with 1-$\sigma$ uncertainties noted by
  errobars. A weak trend at large Y values indicates possible
  effects from charge transfer efficiency.}
\end{figure*}

\begin{figure}
\epsscale{1.25}
\plotone{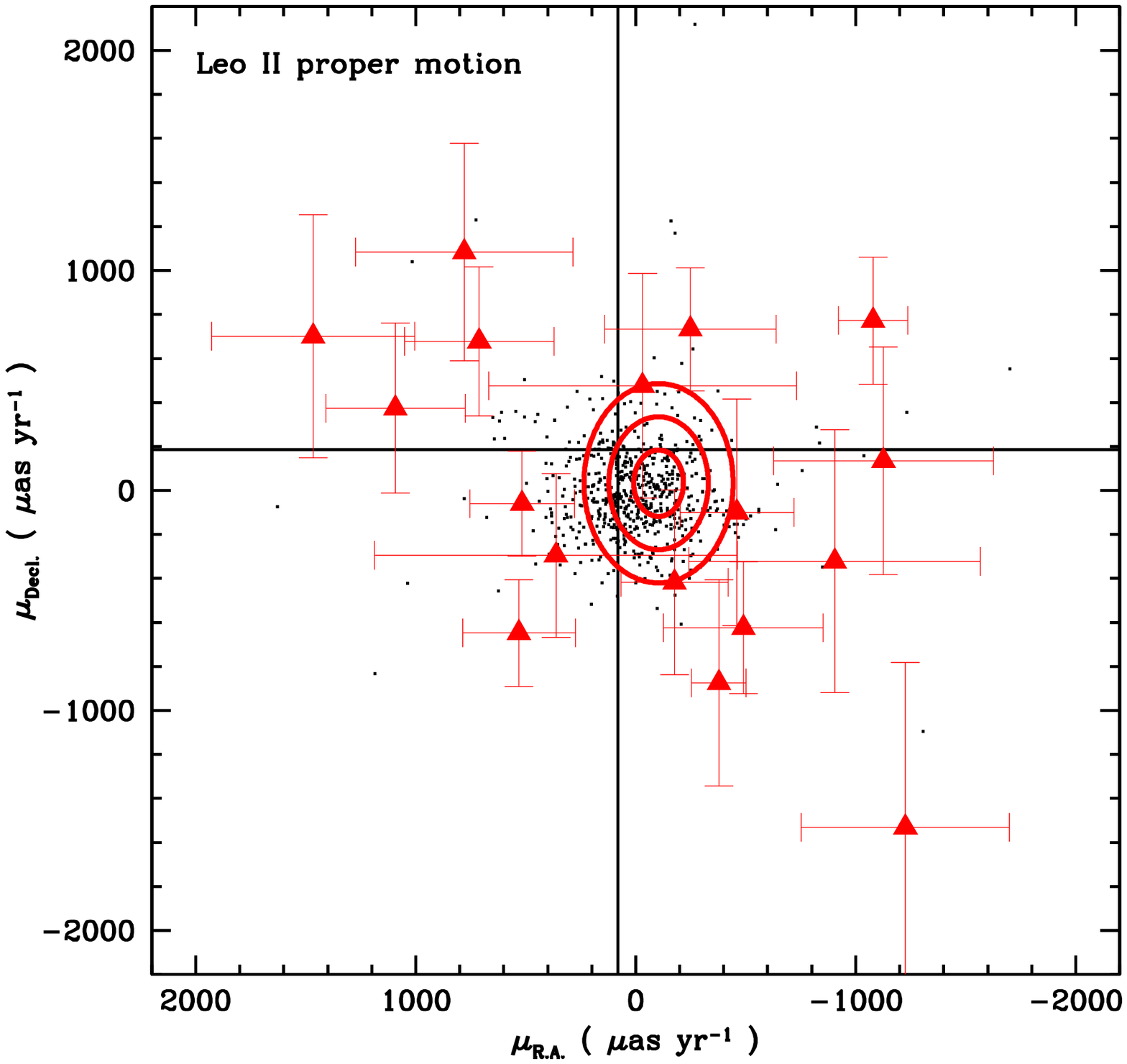}
\caption{Relative proper motion of the sources in the Leo~II
  field. Extragalatic
sources are plotted in red and as filled triangles, with errorbars
showing the estimated 1-sigma measurement errors. The concentric
circles show the 1$\sigma$, 2$\sigma$, and 3$\sigma$ confidence
limits for the systemic proper motion of the extragalctic sources in
the Leo~II reference frame. The crosshairs are centered on the expected
motion of the extragalactic sources if Leo~II were a static target
(i.e., the expected proper motion caused by the reflex motion of the
Sun in the extragalactic rest frame).}
\end{figure}

The identification of a set of extra-galactic sources makes it
possible, in principle, to estimate a proper motion for Leo~II by
measuring the systemic proper motion of these extra-galactic sources
relative to the astrometric reference frame defined by the stellar
members of Leo~II. The {\em absolute} proper motion of Leo~II with
respect to the extragalactic rest frame is then simply the vector
opposite to this {\em relative} proper motion of extragalactic
sources. As this represents the proper motion measured from the
heliocentric rest frame, the component of the solar motion around the
Galaxy must the be taken into account to determine an absolute proper
motion relative to a reference frame at rest relative to the Galactic
barycenter, from which the orbit of Leo~II can be calculated.

First we examine the proper motion in the direction of the X and Y
pixel positions ($\mu_X$, $\mu_Y$). These are different from R.A. and
Dec., because each of the three WF2, WF3, and WF4 cameras has its X-Y
axis oriented differently. The X and Y positional accuracy is thus
more sensitive to instrumental effects (systematic errors in
particular), and better sets the astrometric accuracy of our
measurements. The proper motion distribution is plotted in Fig.~5
(left panels) as a function of magnitude. Sources brighter that
F814W=24.0 --- a total of 634 objects --- have mean values
$(\bar{\mu_X},\bar{\mu_Y})=(-0.23,-0.08)$ millipixels yr$^{-1}$ with a
dispersion $(\sigma_{\mu_X},\sigma_{\mu_Y})=(2.54,2.60)$ millipixels
yr$^{-1}$. The low mean values are consistent with these brighter
sources being part of the astrometric reference grid (of which they
are a subset). The dispersion values yield an estimate of the
intrinsic uncertainty in the proper motion measurement for individual
point sources, which is thus on the order of $\pm0.0025$ pixels
yr$^{-1}$ along both lines (X) and columns (Y) on the WFPC2
frames. The fact that the scatter values are very nearly identical in
X and Y suggests that charge transfer efficiency (CTE) effects are
negligible in this context, otherwise one would expect a larger
scatter in the proper motion along the columns (Y). Sources fainter that
F814W=24.0 --- 2,593 sources --- have mean proper motion values
$(\bar{\mu_X},\bar{\mu_Y})=(+0.20,+0.01)$ millipixels yr$^{-1}$ with a
dispersion $(\sigma_{\mu_X},\sigma_{\mu_Y})=(6.30,6.10)$ millipixels
yr$^{-1}$. The astrometric errors are thus a factor 2.5 larger than for
the brighter sources; the scatter is also observed to increase as the
magnitudes get fainter.

To further examine possible CTE effects, we plot $\mu_X$ a a function
of $X$ and $\mu_Y$ as a function of $Y$ (Fig.~5, right panels). CTE
effects yield systematic offsets between the centroid of bright and faint
sources at large $Y$. Degradation of CTE over time would increase this
offset in the later epochs, resulting in an apparent proper motion
$\mu_Y$ of the faint sources relative to bright ones. In Fig.~5, we
plot $\mu_X$,$\mu_Y$ as individual points for the bright sources
(F814W$<$24.0), and plot the mean proper motions
$\bar{\mu_X},\bar{\mu_Y}$ for the faint sources, in 7 bins along
the X and Y positions (broken lines). For the faint sources, the
dispersion $\sigma_{\mu_X},\sigma_{\mu_Y}$ is noted for each bin with
errobars. We observe that the bright sources do not show any systemic
offset in their proper motion as a function of X or Y, except for
perhaps a weak increase in $\mu_Y$ for $Y>600$. The faint
sources have their mean proper motions $\bar{\mu_X},\bar{\mu_Y}$
within 3$\sigma$ of 0.0 for all X and Y; there is, however, a weak trend
of $\bar{\mu_X}$ increasing with X, and $\bar{\mu_Y}$ increasing with
Y, which could possibly be due to CTE effects. However, the
extra-galactic objects are also bright sources, and we expect them to
behave more like the bright stars, which show negligible CTE effects,
if any.

The 17 extra-galactic (resolved) sources have mean motion
values $(\bar{\mu_X},\bar{\mu_Y})=(-0.18,-1.37)$ millipixels yr$^{-1}$
with a dispersion $(\sigma_{\mu_X},\sigma_{\mu_Y})=(6.71,8.04)$ millipixels
yr$^{-1}$. Their scatter is thus significantly larger than that of the
stellar (resolved) sources. Additionally the scatter appears to be
larger in Y, though this may be due to small number statistics. In any
case, the scatter in the motions is clearly larger for the extragalactic
sources by a factor of about 3. Part of this scatter could be due to a
net relative proper motion of the extra-galactic sources: this is
because the WF2, WF3, and WF4 frames have their XY grids rotated by 90
degrees; a net offset in ($\mu_{\alpha}$,$\mu_{\delta}$) would result
in a scatter in ($\mu_{X}$,$\mu_{Y}$), since the sources come from
different frames. Indeed, the opposite should be true for systematic
offsets in ($\mu_{X}$,$\mu_{Y}$), which would conspire to yield a
large scatter in ($\mu_{\alpha}$,$\mu_{\delta}$); see below. Beside the
possibility of a net proper motion, the larger scatter in the
extra-galactic sources could simply be due to their resolved
nature. The ePSF would not be the best model for those sources, which
would result in larger astrometric errors. Individual ePSF profiles,
tailor-made for each source, would yield more accurate results as
suggested by \citet{Mahmud2008}; however such models require large
numbers of sampling ($>100$) of {\em each object individually},
whereas the ePSF of point sources is built from the sample of all of 
the point sources on the image, requiring fewer exposure frames. Our
current dataset does not contain nearly enough frames of Leo-II to
build such individual ePSF profiles.

Reported in the equatorial frame, the stellar sources in Leo~II with
F814W$<$24 mag have mean proper motion
($<\mu_{\alpha}>$,$<\mu_{\delta}>$)=(-2.7,-7.8) $\mu$as yr$^{-1}$ with a
dispersion ($\sigma{\mu_{\alpha}}$,$\sigma{\mu_{\delta}}$)=(271.9,297.2)
$\mu$as yr$^{-1}$. The dispersion yields an estimate of the accuracy
in the proper motion of invidual sources, while the near-zero value of
the mean proper motion confirms the stability of our astrometric
reference frame. The distribution of proper motions for these
unresolved sources is shown in Fig.6 (dots). 

The proper motion distribution of our 17 selected extra-galactic
sources is also plotted in Fig.6, where we also plot the estimated
internal uncertainties in the proper motion estimates for each individual
source (errorbars). The unweighted mean value is
($\bar{\mu_{\alpha}}$,$\bar{\mu_{\delta}}$)=($+$31.9,$+$25.9) $\mu$as
yr$^{-1}$ and the points have a dispersion
($\sigma{\mu_{\alpha}}$,$\sigma{\mu_{\delta}}$)=(790.5,683.9) $\mu$as
yr$^{-1}$. Since one pixel on the WF cameras is $\approx100$mas, a
700$\mu$as yr$^{-1}$ motion is equivalent to 7 millipixels
yr$^{-1}$. Hence the scatter in ($\mu_{\alpha}$,$\mu_{\delta}$) is
comparable to the scatter in ($\mu_{X}$,$\mu_{Y}$). However, the
scatter in the data points is significantly larger than the estimated
statistical uncertainties of the individual points, which are $\pm450$
$\mu$as yr$^{-1}$ on average. This indicates that the proper motion of
the extra-galactic sources suffers from systematic errors, which are
probably due to the sources being resolved and the local ePSF a
marginal fit to their profile. The dispersion in the data points
suggests that the uncertainties on individual measurements are
underestimate by a factor $\approx1.7$. We account for these
systematic errors by increasing the uncertainties of the individual
proper motions by the same factor. We then estimate the systemic
proper motion of the extragalactic sources by calculating their weighted
average. We find the systemic proper motion to be
($<\mu_{\alpha}>$,$<\mu_{\delta}>$)=($-$104$\pm$113,+33$\pm$151)
$\mu$as yr$^{-1}$, where the quoted errors are the 1-$\sigma$
uncertainties. The 1-$\sigma$, 2-$\sigma$, and 3-$\sigma$ limits are
plotted as ellipses in Fig.6. 

The proper motion of Leo~II in the extra-galactic frame is the
opposite of this vector. We therefore estimate that Leo~II has an
absolute proper motion
($<\mu_{\alpha}>$,$<\mu_{\delta}>$)=($+$104$\pm$113,$-$33$\pm$151)
$\mu$as yr$^{-1}$, which is the proper motion as observed from the
heliocentric rest frame. If Leo~II had no net motion in the plane of
the sky, we would expect the dwarf spheroidal to have a net proper
motion due to the reflex motion of the Sun in the Galactic rest frame;
based on the distance to Leo~II, this reflex motion is predicted to
yield a relative proper motion
($\mu_{\alpha}$,$\mu_{\delta}$)$_{reflex}$=(+80,+186) $\mu$as
yr$^{-1}$. This value is denoted by the crosshairs in
Fig.6, and must be accounted for in calculating the orbital motion of
Leo~II. Adding up this component due to the reflex motion of the Sun
yields for Leo~II a proper motion
($\mu_{\alpha{0}}$,$\mu_{\delta{0}}$)=($+184\pm113$, $+153\pm151$)
$\mu$as yr$^{-1}$ in the {\em Galactic rest frame}. This indicates
that the transverse motion of Leo~II is detected at the 2-$\sigma$
confidence level.

\section{Space velocity and the orbit of Leo~II}

To evaluate the space motion of Leo~II and, in particular, to constrain
the parameters of its orbit, it is useful to report the position and
motion of Leo~II in a reference frame centered on, and at rest with
respect to, the Galactic center. The most straightforward method is to
start with the angular position ($l,b$) and proper motion ($\mu_l,\mu_b$) in the
galactic coordinate system, and include both the measured distance
from the Sun ($r=233\pm15$ kpc) and heliocentric radial velocity ($v_r
= +79$ km s$^{-1}$) of Leo~II. The position ($r,l,b$) and motion
($v_r,\mu_l,\mu_b$) in this spherical, right-handed coordinate system
have a Sun-centered, Cartesian equivalent ($X,Y,Z$) and ($U,V,W$)
defined as:
\begin{eqnarray}
X &=& r\ \cos{l} \cos{b} \\
Y &=& r\ \sin{l} \cos{b} \\
Z &=& r\ \sin{b} \\
U &=& -4.74\ r\ ( \mu_l \sin{l}\ + \mu_b\ \cos{l} \sin{b}\ ) + v_r\ \cos{l} \cos{b} \\
V &=& \phantom{-}4.74\ r\ ( \mu_l \cos{l}\ - \mu_b\ \sin{l} \sin{b}\ ) + v_r\ \sin{l} \cos{b} \\
W &=& \phantom{-}4.74\ r\ \mu_b \cos{b}\ + v_r\ \sin{b}
\end{eqnarray}
with X, Y, and Z expressed in kpc, and U, V, W expressed in km
s$^{-1}$.

Here we recalculate the angular position and distance of Leo~II in a
galactic coordinate system defined by a Cartesian set
($X_G,Y_G,Z_G$) and ($U_G,V_G,W_G$) such that the system has its
origin at the Galactic center and, by definition, is also at rest with
respect to the Galactic center. Positions and motions can be
transformed to this system from the Sun-centered positions and motions
following:
\begin{eqnarray}
X_G &=& X-7.94 \\
Y_G &=& Y \\
Z_G &=& Z \\
U_G &=& U+10.00 \\
V_G &=& V+5.25+225.0 \\
W_G &=& W+7.17
\end{eqnarray}
where we have adopted the galactocentric distance of 7.94 kpc for the
Sun from \citet{GUB2008},
a relative motion of the Sun to the local standard of rest
(U$_{\odot}$,V$_{\odot}$,W$_{\odot}$)=(+10.00, +5.25, +7.17) km
s$^{-1}$ \citep{1998MNRAS.298..387D}, and a Galactic rotation velocity
of 225 km s$^{-1}$ at the solar vicinity. In this system, we calculate
for Leo~II a position ($X_G,Y_G,Z_G$)$\approx$($-$76.8, $-$58.2,
214.8) kpc. We convert the motion into this Galactic reference frame,
using a Monte Carlo simulation to evalutate the range of
uncertainties, and find ($U_G,V_G,W_G$)=(100.8$\pm$126.6, 216.0$\pm$156.9,
118.1$\pm$49.3) km s$^{-1}$, for a total Galactocentric space
velocity of $v_{\rm GRF} = 266.1\pm$128.7 km s$^{-1}$.


This Galactocentric Cartesian system also has an associated spherical
coordinate system ($r_G,l_G,b_G$) and
($v_{r_G},\mu_{b_G},\mu_{b_{G}}$), which are related to the Cartesian
system ($X_G,Y_G,Z_G$) and ($U_G,V_G,W_G$) with the same relationships
as in Eqs. 3--8. In this system, we calculate that Leo~II
is located at
($r_G,l_G,b_G$)$\approx$(235.5 kpc, 217.132$^{\circ}$, +65.839$^{\circ}$). Its
Galactocentric radial velocity and proper motion is
($v_{r_G},\mu_{l_G},\mu_{b_{G}}$) = (21.5$\pm$4.3 km
s$^{-1}$, $-$99.8$\pm$147.8$\mu$as yr$^{-1}$,
215.6$\pm$112.8$\mu$as yr$^{-1}$). This proper motion yields a total
transverse velocity
$v_{t_G}$=265.2$\pm$129.4 km s$^{-1}$, a large value compared to
the estimated radial motion $v_{r_G}$=21.5$\pm$4.3 km s$^{-1}$. This
yields a mostly tangential space velocity,
from which we infer that Leo~II either has a low-eccentricity orbit,
or is currently close to perigalacticon or apogalacticon distance. 

%

%
%
%
%

With these kinematic components at hand  we integrated its orbit for a
variety of Galactic potentials,  where the integration was carried out
14 Gyr backwards and forward in time. All potentials used for our test
cases have disk and bulge components
\citep[e.g.][]{1990ApJ...348..485P,1991RMxAA..22..255A,1998MNRAS.294..429D},
which, however,  do not contribute significantly to the potential
given the large distance (in particular its large height above the
disk) of the dSph. More importantly, this large distance renders
remote satellites like Leo~II important tracers of the Galactic halo
and its total mass distribution \citep{1999MNRAS.310..645W}.  This is
already obvious from plotting the escape velocity of such models at
the position of Leo~II (Fig.~7).

\begin{figure}
\includegraphics[width=1\hsize]{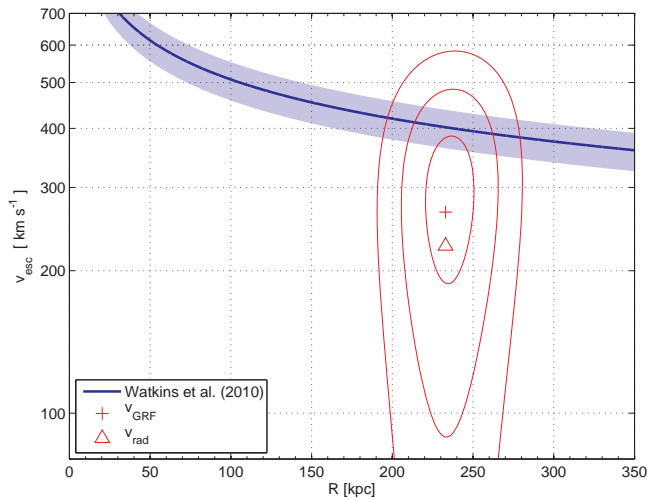}
\caption{Escape velocity as a function of Galactocentric distance for
  the best-fit halo model from Watkins et al. (2010), which
  adopts  a total mass out to 300 kpc of $2.7\times10^{12}
  M_{\odot}$. Both under exclusion (triangle) or inclusion (cross) of
  our proper motion measurement it appears likely that  Leo~II is
  bound to the MW {\em at present}. Also shown are the 1, 2, and
  3$\sigma$- contours).}
\end{figure}

In this figure, we show the best-fit scale-free potential
from \citet{2010MNRAS.406..264W} with its total mass of
$(2.7\pm0.5)\times10^{12} M_{\odot}$ that has been established using
tracer satellites out to larger distances of $\sim$300 kpc, but we
note that the same arguments hold for the various (dark) MW halo
models found in the literature to date
\citep[e.g.][]{2005MNRAS.364..433B,Dehnen2005,Dehnen2006}. Even if our
proper motion measurements were significantly in error and assuming
that only the radial velocity component contributed reliably, Leo~II's
space velocity of  v$_{\rm  GRF,\,rad}=266\pm128$ km\,s$^{-1}$ is still
high and similar to the value we obtain from using the full kinematic
information. Thus, it is the large distance and radial velocity
which govern its derived dynamics. The comparison in Fig~7 implies that
Leo~II is a true, bound satellite to the MW: the local escape velocity
exceeds the space velocity of Leo~II by 1.0$\sigma$. 
%
We note, however, that the concept of ``escape velocity'' for the
present arguments should be taken with caution, since its definition
commonly neglects all mass {\em outside} of the considered
radius. Moreover, recent evidence suggests that the rotation velocity
of the Milky Way may be larger than used in the present work and all
current Galactic potential models, at $\Theta_0\sim 250$ km\,s$^{-1}$
\citep{2009ApJ...700..137R}. We will address these issues in a future work
dealing with the implications of Leo II's boundedness for current MW mass
models (Koch et al. in prep.). 

 
In Fig.~8 we show the orbital solution based on the aforementioned
integrations and our fiducial proper motion measurements. While
currently formally ``bound'' to the MW potential, it is evident that
Leo~II has come a long way over its recent past. Currently at
pericenter, its ``orbital'' period implied from the calculations is 50
Gyr and, rather than an apocenter, it reaches its largest distance
from the MW at the limits of our integrations, at $\sim$1.8
Mpc. Considering these time scales and large distances from the actual
Galactic potential, any timing argument and orbital solution becomes
necessarily unreliable and one would need to account for the entire
dynamical history of the Local Group
\citep{1989ApJ...345..108P,1994AJ....107.2055B,2007A&A...463..427P}.

\begin{figure}
\includegraphics[width=1\hsize]{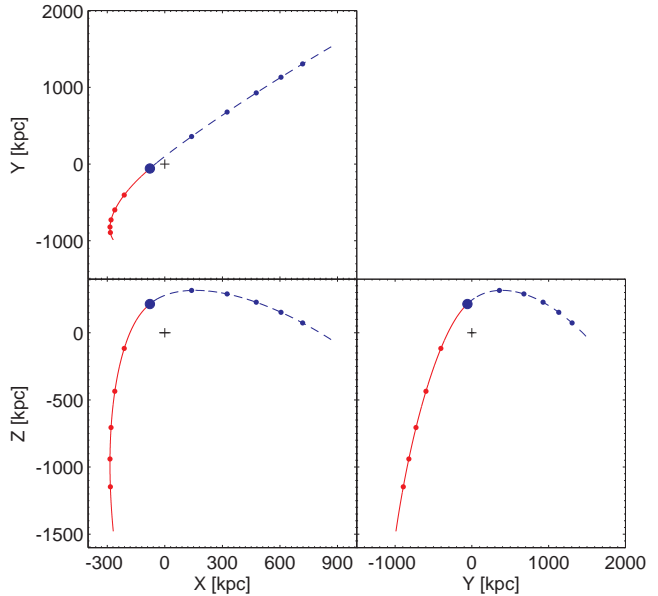}
\caption{Orbital solution from our best velocity measurements in
  Galactic coordinates. Red and blue curves refer to backwards and
  forward integration, respectively. Its current location (blue solid
  circle)  and the position of the MW (cross) are also indicated, as
  well as its position along the orbit at $\pm$(2, 4, 6, 8, 10) Gyr
  (small points). }
\end{figure}

In order to better understand Leo~II's orbital characteristics, we
recomputed its parameters in a Monte Carlo sense by varying position,
velocity, and proper motion within their uncertainties. As a result,
we find periods shorter than 6.5 (9.1, 11.6) Gyr in only 5\% (10\%,
15\%) of the realizations. Likewise, in only 4\% (6\%, 12\%) of the
cases can we reproduce pericenters within 100 (150, 200) kpc, thus
bringing Leo~II into the realm of the majority of the MW's closer,
bound satellites \citep{2003AJ....125.1926G,2009AN....330..675K}.

Therefore, we conclude that this dSph is rather an isolated Local
Group satellite that is falling into the MW regions and passing its
(dark) halo for the first time (e.g., Chapman et al. 2007; Majewski et
al. 2007). In fact, interactions or common origins of dynamical
interlopers like Leo~II with other Local Group systems appears
plausible \citep{2007MNRAS.379.1475S}.

\section{Conclusions}

Our astrometric analysis provides the first-time detection of a proper
motion for the dwarf satellite galaxy Leo~II. The transverse motion of
Leo~II is detected, but only after accounting for the solar reflex
motion.

The question of its origin and whether or not Leo~II is a bound
satellite has important implications for our understanding of the
infall and merging of cosmological subhalos and the dynamics and
structure of the Local Group. On the other hand, Watkins et
al. (2010; their Fig.~5) estimate that the cumulative contribution of
Leo~II to the mass budget of the MW is typically less than 8\% (thus
$< \pm  0.2\times10^{12} M_{\odot}$). Removing it from the family of
bound MW satellite tracers would thus only have a minor impact.


We can compare the idea that Leo~II spent most of its life in
seclusion from the MW to the properties of other isolated Local Group
dSphs. \citet{1996AJ....111..777M}, \citet{2002MNRAS.332...91D} and
\citet{2007AJ....133..270K} find evidence of star
formation as recent as 2 Gyr ago, which is comparable to another,
remote MW dSph, Leo~I \citep{1999AJ....118.2245G}. Tucana (at 900 kpc)
and Cetus (775 kpc from the MW) are associated with neither M31 nor the MW
and neither of them shows any evidence of past or present interactions
with these massive galaxies
\citep{2007MNRAS.375.1364L,2009A&A...499..121F}. Despite the lack of
any obvious, dynamic gas removing
agents at those large distances, none of these dSphs contains any gas;
in particular Tucana appears to have had no active star formation over
the past 8--10 Gyr \citep{1996A&A...315...40S}.  This may imply that
galaxies like Leo~II  had experienced efficient galactic winds, which
is also consistent with its flat age-metallicity relation during the
first seven or so Gyr after the Big Bang \citep{2007AJ....133..270K}.

\acknowledgements

We thank L. Watkins, M.Irwin, and M. Wilkinson for helpful
discussions and W. Dehnen for providing his potential code. Support
for program \#11182 was provided by NASA through a grant from the
Space Telescope Science Institute, which is operated by the
Association of Universities for Research in Astronomy, Inc., under
NASA contract NAS 5-26555. SL also acknowledges support from National
Science Foundation grant AST-0607757 and from NASA STScI grant
AR-11770. AK acknowledges support by an STFC postdoctoral
fellowship and funding by the DFG through Emmy-Noether grant Ko
4161/1. RMR acknowledges support from NASA STScI grants GO-9817,
GO-11182, and AR-11770.


\end{document}